\documentstyle[12pt]{article}
\begin{document}
{\large Cosmological density perturbation in two fluid models on spin dominated inflation and the stability of Friedmann metric }
\vspace{2cm}
\noindent
\begin{center}
{\large L. C.
Garcia de Andrade\footnote{Departamento de Fisica Teorica-Instituto de F\'{\i}sica -UERJ, Rua S\~{a}o
Francisco Xavier 524, Maracan\~{a},R.J
Cep: 20550 - Rio de Janeiro, Brasil.E-mail:garcia@dft.if.uerj.br}}
\end{center}
\vspace{2.0cm}
\begin{abstract}
A spinning fluid embedded in a space section flat Friedmann model is used to compute the cosmological density perturbation of the model.The spinning fluid obeys the Einstein-Cartan field equations while the Friedmann embedded model is assumed to obey the general the Friedmann equation in General Relativity.It is shown that at the early epochs of the Universe the spin-torsion contribution to the density perturbation on the matter phase dominates over the radiation perturbation.On the other hand at the present epochs of the  Universe the spin-torsion contribution is more efficient in the matter perturbation.Stability of the Friedmann solution in Einstein-Cartan cosmology is also discussed. 
\end{abstract}
\vspace{1.0cm}
\begin{center}
\end{center}
\newpage
\pagestyle{myheadings}
\markright{\underline{Cosmological density perturbations stability of Friedmann metric.}}
Recently D.Palle \cite{1} computed the evolution equatioon of density perturbation for a Bianchi type III model in Einstein-Cartan cosmology.More recently a dilaton model for the de Sitter inflation in EC gravity was proposed where the spin-torsion density fluctuation has been computed \cite{2} and the evolution equation computed \cite{3}.It is well know in standard general relativistic cosmology that specially in the particular case of Newtonian cosmology simple models can be designed to obtain the density perturbations without solving the evolution equation \cite{4}.In this note we propose a two fluid very simple model in Einstein-Cartan gravity to compute the cosmological density perturbations without solving the evolution equation of density perturbations.In the absence of spin our results reduce to the ones in General Relativity \cite{5}.Dominance of the spin-torsion density in the early epochs of the Universe occurs in the case of density perturbations for the matter phase while at the present epoch spin-torsion stronger contribution is to the density perturbations is on radiation era.This fact seems to be physically explainable since the early epochs of the Universe are carachterize by the fact that the matter-radiation coupling is much stronger than in the present day Universe.Let us start by considering the Friedmann metric 
\begin{equation}
ds^{2}=dt^{2}-a^{2}(t)(\frac{dr^{2}}{1-kr^{2}}+r^{2}(d{\theta}^{2}+sin^{2}{\theta}d{\phi}^{2}))
\label{1}
\end{equation}
where $a(t)$ is connected to the Hubble parameter H by the relation $H=\frac{\dot{a}}{a}$ and k is the spatial curvature constant which is $k=-1,0,+1$ according the 3-space is open,flat or closed respectively.Considering a spherical region of radius ${\lambda}>d_{H}$ which is the Hubble radius containing spinning matter with mean density ${\rho}_{1}$,embedded in a $k=0$ Friedmann universe ${\rho}_{0}$ where ${\rho}_{1}={\rho}_{0}+{\delta}{\rho}$, ${\delta}{\rho}>0$ and represents a small density perturbation.As point it out by Padmanabhan \cite{5} the spherical symmetry implies that the inner region is not affected by the matter outside.Thus the inner region evolves as a $k=+1$ Friedmann Universe .The two regions form a kind of two fluid the inner fluid being a spinning fluid obtained from the perturbation.Since the inner fluid is a spin fluid it may obey the Einstein-Cartan gravity with the following form
\begin{equation}
{H_{1}}^{2}+\frac{1}{a_{1}}=\frac{8{\pi}G}{3}({\rho}_{1}-2{\pi}G{{\sigma}^{2}})
\label{2}
\end{equation}
where ${\sigma}^{2}$ represents the averaged squared value of the spin-torsion tensor.The outer fluid obey the GR equation
\begin{equation}
{H_{0}}^{2}=\frac{8{\pi}G}{3}({\rho}_{0})
\label{3}
\end{equation}
where both universes are compared, the perturbed and the background universe when their expansion rates are equal are equal,or we compare their densities at the time t when $H_{1}=H_{0}$.Since
\begin{equation}
\frac{{\delta}{\rho}}{{\rho}_{0}}=\frac{{\rho}_{1}-{\rho}_{0}}{{\rho}_{0}}
\label{4}
\end{equation}
Performing the above substitution we obtain
\begin{equation}
\frac{{\delta}{\rho}}{{\rho}_{0}}=\frac{1}{{8{\pi}G}{a^{2}}_{1}}+\frac{2{\pi}G{\sigma}^{2}}{{\rho}_{0}}
\label{5}
\end{equation}
To compute this equation in terms of the cosmic expansion $a$ we simply remember that the spin-torsion density is proportional to $\frac{{S_{0}}^{2}}{a^{6}}$.where ${S_{0}}^{2}=n^{2}h^{2}$ is the spin tensor squared and n is equal to the number of nucleons in the Universe and h is the Planck constant.Substituting this value into the equation (\ref{5}) and considering that $a_{1}$ is approximatly equal to $a_{0}$.This permitt us to see how the $\frac{{\delta}{\rho}}{{\rho}_{0}}$ scales with a.Since in the radiation dominated era ${\rho}_{0}{\alpha}a^{-4}$ and ${\rho}_{0}{\alpha}a^{-3}$ in the matter dominated phase, substitution of these  values into the equation (\ref{5}) yields respectively
\begin{equation}
\frac{{\delta}{\rho}}{{\rho}_{0}}|_{R}=\frac{a^{2}}{8{\pi}G}+{\beta}G{S_{0}}^{2}{a}^{-2}
\label{6}
\end{equation}
and
\begin{equation}
\frac{{\delta}{\rho}}{{\rho}_{0}}|_{M}=\frac{a}{8{\pi}G}+{\beta}G{S^{2}}_{0}{a}^{-3}
\label{7}
\end{equation}
others zero.Here ${\beta}$ is a constant.To write down these expressions in terms of cosmic time to be able to check the observational results with our computations we need to solve equation (\ref{3}) in terms of the cosmic scale factor a in terms of time.Solution of this equation in terms of time yields the following results $a|_{R}{\alpha}t^{\frac{1}{2}}$ and $a|_{M}{\alpha}t^{\frac{2}{3}}$.Substitution of these values into the expressions (\ref{6}) and (\ref{7}) we obtain respectively
\begin{equation}
\frac{{\delta}{\rho}}{{\rho}_{0}}|_{R}=\frac{t}{8{\pi}G}+{\beta}G{S_{0}}^{2}{t}^{-1}
\label{8}
\end{equation}
and
\begin{equation}
\frac{{\delta}{\rho}}{{\rho}_{0}}|_{M}=\frac{t^{\frac{2}{3}}}{8{\pi}G}+{\beta}G{S_{0}}^{2}{t}^{-2}
\label{9}
\end{equation}
Notice therefore that in the case of early epochs in the Universe where $t-0$ the spin-torsion effects appear to be stronger on matter perturbation contributing to the decoupling of matter and radiation this is explained since the torsion acts only on fermions and do not clearly interacts with radiation.Nevertheless in the present epoch the spin-torsion decays very quick which is the reason the detection of torsion is difficult froom the cosmological point of  view nowadays.Now we tuern to the problem of the stability of the Friedmann metric in Einstein-Cartan cosmology.Nurgaliev and Ponomariev \cite{6} investigated the early evolutionary stages of the Universe in the realm of Einstein-Cartan cosmology by considering an ideal fluid of nonpolarised fermions.They show that for this particular case the Friedmann solution was stable to small homogeneous and isotropic perturbations.They also conclude that the increase in entropy could lead to the evolution of the initially small stable oscillations into large ones.The Universe would begin after a definite number of oscillations.Calculations of small perturbations in their model would shown some instability which would prepare for the necessary initial conditions for the growth of pertubations in the nonlinear regime.Now we make use of their result to olace a lower limit to the spin-torsion primordial density fluctuation obtained from the Einstein-Cartan gravity and COBE satellite data \cite{1,2}.In this way structure formation like Galaxies formation would not suffer a great influence from torsion \cite{4}.It is also shown that at early stages of the Universe Bianchi type III models with expansion and rotation may not depend at all from torsion.Other Bianchi types like an oscillating Bianchi type IX model in Einstein-Cartan \cite{7} gravity have also been recently investigated.
\begin{equation}
ds^{2}=dt^{2}-a^{2}(t)(dz^{2}+dx^{2}+dy^{2})
\label{1}
\end{equation}
The Friedmann equation thus becomes
\begin{equation}
\frac{{\ddot{a}}}{{a}}=\frac{4{\pi}G}{3}({\rho}-8{\pi}G{\sigma}^{2})
\label{2}
\end{equation}
Making an homogeneous and isotropic small perturbation on the Friedmann yields
\begin{equation}
\frac{{\delta}{\ddot{a}}}{{\delta}{a}}=-\frac{4{\pi}G}{3}({\rho}-8{\pi}G{\sigma}^{2})-\frac{4{\pi}G}{3}(\frac{{\delta}{\rho}}{{\delta}a}-8{\pi}G\frac{{\delta}{\sigma}^{2}}{{\delta}{a}})
\label{3}
\end{equation}
Substitution of the well-known relation
\begin{equation}
\frac{{\delta}{\rho}}{{\delta}a}=-3\frac{\rho}{a}
\label{4}
\end{equation}
we obtain
\begin{equation}
\frac{{\delta}{\ddot{a}}}{{\delta}{a}}=-\frac{4{\pi}G}{3}({\rho}-8{\pi}G{\sigma}^{2})+\frac{4{\pi}G}{3}(\frac{{\rho}}{a}+8{\pi}G\frac{{\delta}{\sigma}^{2}}{{\delta}{a}})
\label{5}
\end{equation}
and the stability condition $\frac{{\delta}{\ddot{a}}}{{\delta}{a}}<0$ implies 
\begin{equation}
(1-8{\pi}G\frac{{\delta}{\sigma}^{2}}{{\delta}{\rho}})<0
\label{6}  
\end{equation}
Since the matter density ${\rho}>0$ this implies the following condition
\begin{equation}
{{\delta}{\sigma}^{2}}>\frac{1}{8{\pi}G}\frac{{\delta}{\rho}}{{\rho}_{0}}{\rho}_{0}
\label{7}  
\end{equation}
where we take${\rho}_{0}=10^{-31}g{cm}^{-3}$ as the matter density of the Universe and from the COBE data $\frac{{\delta}{\rho}}{{\rho}_{0}}=10^{-5}$.From these data formula (\ref{7}) yields the following lower limit for the spin-torsion fluctuation as
\begin{equation}
{\delta}{\sigma}^{2}>10^{-28} cgs units
\label{8}
\end{equation}
 this result was expected since the spin-torsion density decreases with the expansion and is redshifted with inflation.This conjecture has been proposed recently by Ramos and myself \cite{8}.As pointed out by Nurgaliev and Ponomariev \cite{6} the increase in the entropy may trigger the growth in the inhomogeneities.There is no compelling reason to believe that this would not happen here.Moreover Nurgaliev and Piskareva \cite{9} have also investigate the structural stability of cosmological models in Einstein-Cartan gravity.A more detailed investigation of the matters discussed here including the Bianchi type IX oscilating solution in Einstein-Cartan cosmology may appear elsewhere.
\section*{Acknowledgement}
I am very much indebt to Professors P.S.Letelier,I.Shapiro and my colleague Rudnei de Oliveira for helpful discussions on the subject of this paper.Thanks are also due to an unknown referee for useful comments.Financial support from CNPq. and UERJ is gratefully acknowledged.

\end{document}